\documentclass[preprint,journal]{vgtc}                     %

\onlineid{0}

\vgtccategory{Research}

\vgtcpapertype{application/design study}

\title{\tool{}:  Automatic Grading of D3 Visualizations}

 \author{%
  Matthew Hull,
  Vivian Pednekar,
  Hannah Murray,
  Nimisha Roy, \\
  Emmanuel Tung,
  Susanta Routray,
  Connor Guerin,
  Justin Chen, \\
  Zijie J. Wang,
  Seongmin Lee,
  Mahdi Roozbahani, and 
  Duen Horng Chau
 }

\authorfooter{
  \item Matthew Hull, Vivian Pednekar, Hannah Murray, Nimisha Roy, Emmanuel Tung, Susanta Routray, Connor Guerin, Justin Chen, Zijie J. Wang, Seongmin Lee, Mahdi Roozbahani, and Duen Horng Chau are with Georgia Institute of Technology. Email: \{matthewhull\,$|$\,vpednekar3\,$|$\,hmurray9\,$|$\,\ nroy9\,$|$\,tunge\,$|$\,sroutray\,$|$\,cguerin6\,$|$\,chen3001\,$|$\,jayw\,$|$\,seongmin\,$|$\,\ mahdir\,$|$\,polo\}@gatech.edu\,.
}
\abstract{%
Manually grading D3 data visualizations is a challenging endeavor, and is especially difficult for large classes with hundreds of students. 
Grading an interactive visualization requires a combination of interactive, quantitative, and qualitative evaluation that are conventionally done manually and are difficult to scale up as the visualization complexity, data size, and number of students increase. 
We present \tool{}, a first-of-its kind automatic grading method for D3 visualizations that scalably and precisely evaluates the data bindings, visual encodings, interactions, and design specifications used in a visualization. 
Our method enhances students' learning experience, enabling them to submit their code frequently and receive rapid feedback to better inform iteration and improvement to their code and visualization design. 
We have successfully deployed our method and auto-graded D3 submissions from more than 4000 students in a visualization course at Georgia Tech, and received positive feedback for expanding its adoption. }

\keywords{Automatic grading, D3 visualization, large class, Selenium, Gradescope grading platform}

\teaser{
  \includegraphics[width=\linewidth]{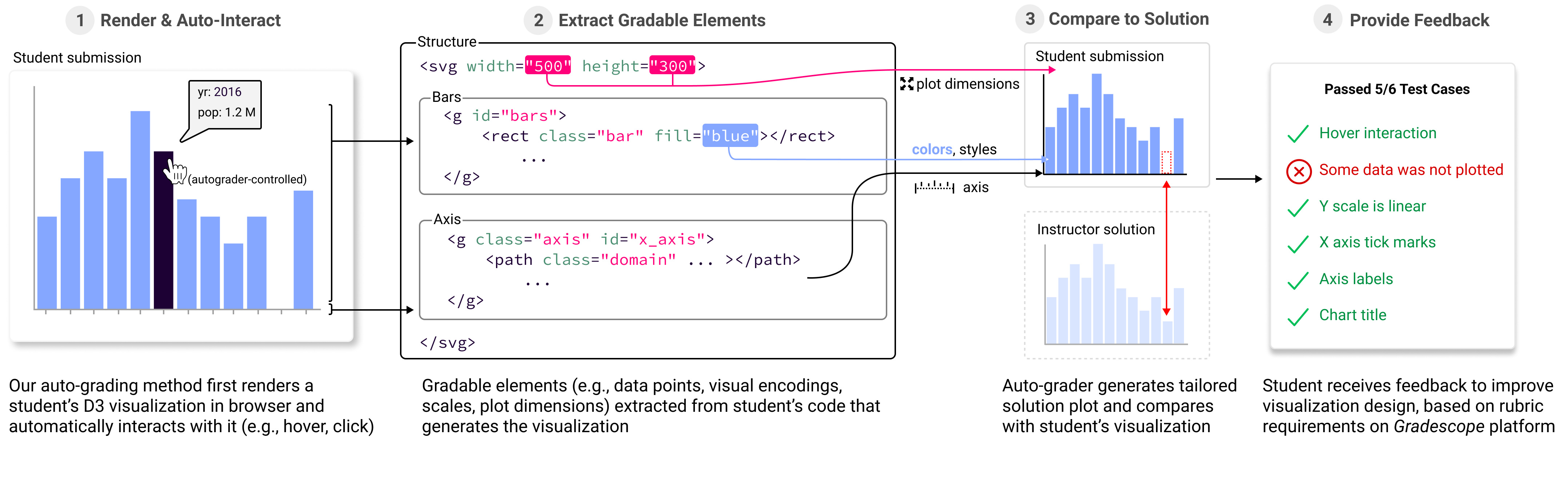}
  \caption{Our auto-grading workflow for D3 visualizations 
  (1) first renders student's plot in a web browser and automatically interacts with it (e.g., hover, click); 
  (2) extracts gradable elements (e.g., data points, visual encodings, plot dimensions) from the plot's backing code; 
  (3) generates a student-specific solution and compares it with student's plot to identify issues; and 
  (4) provides feedback to help student improve their design.}
  \label{fig:teaser}
}

\graphicspath{{figs/}{figures/}{pictures/}{images/}{./}} %

\usepackage{tabu}                      %
\usepackage{booktabs}                  %
\usepackage{lipsum}                    %
\usepackage{mwe}                       %

\usepackage{mathptmx}                  %
\usepackage{xspace}
\usepackage{enumitem}

\usepackage{microtype}                 %
\PassOptionsToPackage{warn}{textcomp}  %
\usepackage{textcomp}                  %
\usepackage{mathptmx}                  %
\usepackage{cite}

\usepackage{enumitem}
\usepackage{tabu}
\usepackage{rotating}
\usepackage{capt-of,etoolbox}
\usepackage{microtype}
\usepackage{bm}
\usepackage{soul}
\usepackage{wrapfig}
\usepackage{graphbox}
\usepackage{mathtools}
\usepackage{balance}
\usepackage[varqu]{zi4}
\usepackage[bb=boondox]{mathalfa}
\usepackage{units}
\usepackage{setspace}
\usepackage{gensymb}
\usepackage{printlen} %
\usepackage{placeins} %
\usepackage{afterpage} %
\usepackage{amsmath}
\usepackage{eqnarray}
\usepackage{bbm}

\newcommand{\todo}[1]{\textcolor{red}{[*** #1 ***]}}
\newcommand{\matthew}[1]{\textcolor{hotpink}{[#1 -matthew]}}
\newcommand{\polo}[1]{\textcolor{purple}{[#1 -polo]}}
\newcommand{\jay}[1]{\textcolor{teal}{[#1 -jay]}}
\newcommand{\seongmin}[1]{\textcolor{agreen}{[#1 -seongmin]}}

\newcommand{\mkclean}{
  \renewcommand{\matthew}[1]{}
  \renewcommand{\jay}[1]{}
  \renewcommand{\seongmin}[1]{}
  \renewcommand{\polo}[1]{}
  \renewcommand{\todo}[1]{}
}
\mkclean{}

\newcommand{\tool}{\textsc{\textsf{VisGrader}}}

\newtoggle{inheader}

\newcommand{\inlinefig}[2]{\protect\includegraphics[align=c, height=#1pt]{figs/#2}}

\begin{document}

\maketitle

\section{Introduction} %
\label{sec:introduction}

\begin{figure*}[!ht]
    \includegraphics[width=\linewidth]{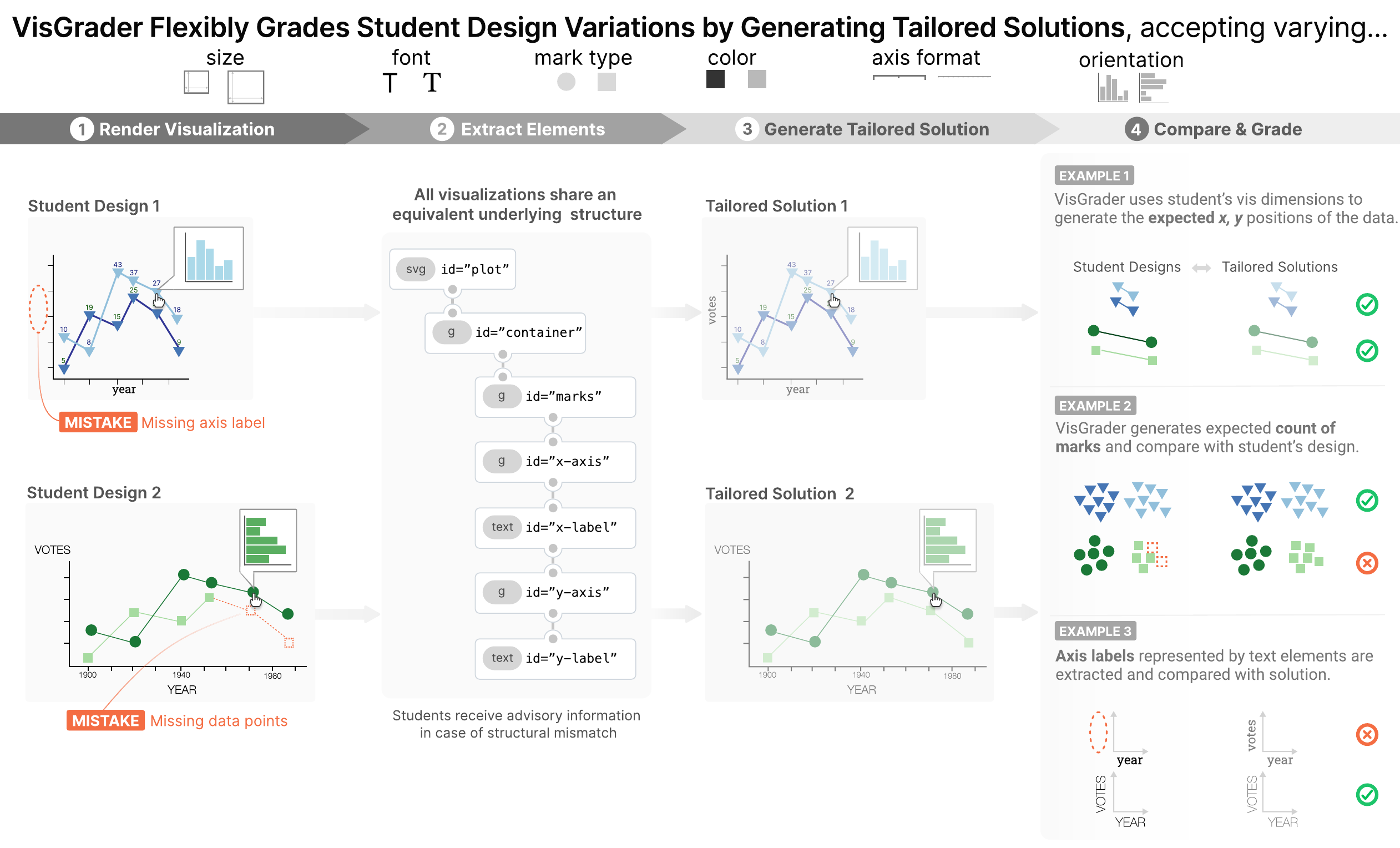}
    \centering
    \caption{\tool{} flexibly evaluates students' design variations. Here, we show two example student designs (on the leftmost) that use different plot sizes, fonts, mark types, colors, axis formats, and bar chart orientations. 
    \tool{} handles these different design variations by
    evaluating their underlying HTML structures.
    \inlinefig{9}{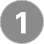} When a visualization is rendered, \inlinefig{9}{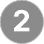} the HTML structure is extracted, and \inlinefig{9}{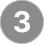} a tailored student-specific solution is generated for \inlinefig{9}{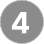} comparison with the student submission. }
    \label{fig:diff_designs}  
\end{figure*}
D3 \cite{2011-d3} is in wide use and remains a highly popular choice in data visualization courses~\cite{CS444DataViz,CS5630DatViz4DS,CSE512DataViz,CS171Viz,CS448DataViz,CS6242DVA}
due to its immense flexibility and foundational stature in the visualization community\cite{Lo2019eLearnVizTools}. 

Manually grading D3 \cite{2011-d3} visualizations is a challenging endeavor, and is especially difficult for large classes with hundreds of students.
Even for a ``simpler'' interactive bar plot visualization, manually grading it usually requires a combination of interactive, quantitative, and qualitative evaluation, with tasks such as verifying the correctness of data bindings and visual encodings, testing interactive elements and actions work as intended (e.g., whether tool-tip displays upon mouse-hover), and determining if a visualization's overall layout and spacing is appropriate.
However, such ``simple'' evaluation tasks quickly become daunting as the visualization complexity and data size increase. For a large class with hundreds or a thousand students, such manual grading becomes extremely tedious \cite{maicus2020autograding}.
When evaluating visualizations on a large scale, it is impractical to exhaustively evaluate and interact with all plotted data.
Furthermore, manual grading presents the logistic challenge of requiring congruence between student and grader environments. 
This necessitates the recreation of the student's environment or in some cases, modification of student code by the grader to run the visualization, resulting in a cumbersome grading process, especially for large classes with many students.
To address the common challenges in grading D3 visualizations, we contribute:
\begin{itemize} [topsep=0mm, itemsep=0mm, parsep=1mm, leftmargin=5mm]
\item{\textbf{\tool{}, the first-of-its-kind automatic grading approach for D3 visualizations}} 
that scalably and precisely evaluates the data bindings, visual encodings, interactions, and design specifications used in a visualization (Fig. \ref{fig:teaser}).
Our method avoids a rigid ``one-size-fits-all'' approach to grading, by offering a novel flexible way that provides students with \textit{tailored feedback} while supporting their design freedom in developing visualizations (Fig. \ref{fig:diff_designs}),
accepting variations in size, font, mark type, color, axis format, orientation and more, provided that students' designs follow the same high-level structures, so that ``gradable'' elements may be extracted for grading (Sec. \ref{sec:method:flexible_grading_design}).
 Our approach utilizes Selenium’s \cite{Selenium} browser automation package to compose rich interaction sequences, known as \textit{Action Chains}\cite{SeleniumActionChain}, that facilitate the evaluation of common user interaction (Sec. \ref{sec:method:eval_interaction}).
As a result, our auto-grader can efficiently and automatically interact with and grade interactions and “hidden” visualization elements that display after a complex sequence of interactions (Fig. \ref{fig:action_chains}). Without our approach, such grading would need to be tediously performed for every student’s submission.
To promote easy adoption of our auto-grading approach by  visualization instructors, 
we have open-sourced it on GitHub at \url{https://github.com/poloclub/visgrader},
which has included
the auto-grader source code,
a series of Jupyter Notebooks that explain how we extract and grade elements, 
an example visualization that demonstrates how auto-grading is applied, 
and a 30-second summary video that provides a high-level overview of the approach and its impacts\footnote{\tool{} video: \url{https://youtu.be/XnTfiAiYLNM}}.

\item{\textbf{Large Scale D3 Auto-Grading}}. 
\tool{} has been used to grade more than 72,000 submissions from more than 4,000 students for four semesters\footnote{Fall 2021, Spring 2022, Fall 2022, Spring 2023} in Georgia Tech's \textit{CSE 6242 Data and Visual Analytics}\footnote{\url{https://poloclub.github.io/\#cse6242}} course\cite{CS6242DVA}.
Each recent semester sees an average of 1,100 students from both undergraduate and graduate programs in both on-campus and online offerings and received positive feedback from both students and instructors.
Our auto-grading approach has enhanced students' learning in multiple ways, enabling them to  receive instant feedback to improve their work, learn best coding practices, and reduce configuration errors.
Our approach has also benefited the instructional staff, such as reducing teaching assistant (TA) grading hours and allowing more time to provide qualitative feedback (Sec. \ref{sec:evaluation}).

\end{itemize}

\section{Background and Related Work}
\label{sec:related_work}

Automated testing tools\cite{Sen2005DARTDA} for software have been in widespread use for many years and derivatives of this methodology have been incorporated into automated grading or ``auto-grading'' of introductory programming assignments\cite{Gulwani2014FeedbackGF,wilcox2015role} in Computer Science courses.

\subsection{Auto-grading Student Programs}
\label{sec:related_work:auto_grade_student}
Typical auto-graders can check for syntax errors\cite{Parihar2017AutomaticGA} and can generate feedback \cite{Gao2016AutomatedFF} that students can access in an effort to improve their implementation and achieve a higher score.
There exists evidence that automated adaptive feedback\cite{Ahmed2020CharacterizingTP} can assist in developing stronger programming competencies in large environments. 
These approaches generally are well-suited toward problems of evaluating a program's ability to successfully execute or its output value and we refer to these as ``value/output checking'' auto-graders. 
An example might be a student writing a simple calculator program in an introductory programming class would implement a set of functions \textit{e.g.,} \texttt{add(x,y)} or \texttt{subtract(x,y)} and the corresponding auto-grader would then call \texttt{add(3, 4)} or \texttt{subtract(10,9)} and check the result.  
A more sophisticated auto-grader might check the output of a machine learning model, e.g., a student could implement a Decision Tree model that the auto-grader will supply some unseen data to ensure its accuracy falls within a certain range.

However, a D3 visualization is decidedly different from a classic computer program in the way that it is intended more for data communication and interaction, requiring a different approach for evaluation. 
Furthermore, even students implementing the same visualization (e.g., a bar chart) may create a great variety of different designs that use different colors, plot sizes, axis formats, and more.  
A trivial ``value-checker'' auto-grader cannot evaluate a submission fairly if it expects only a single color or size.  
If an auto-grader simply checked the x,y coordinates of a plotted datum, it would exist at different positions due to the plot size specified by the student.
Similarly, comparing pixels of rasterized visualizations would lack flexibility for accommodating minor deviations when the submission does not precisely match.

\subsection{Auto-grading Complex Visualization}
\label{sec:related_work:auto_grade_complex}
To the best of our knowledge, we have not seen any work to date that can auto-grade a D3 visualization.
Thus, we discuss related methods for finding and correcting mistakes, and extracting and identifying the underlying structure of a D3 visualization, a critical component for developing automated grading methods.

\subsubsection{Correcting Mistakes in a Visualization}
\label{sec:related_work:auto_grade_complex:correct_mistakes}
Currently, there are some tools and frameworks that can automatically generate or evaluate visualizations that are built for Vega-Lite\cite{2017-vega-lite}, a declarative visualization framework built on D3.  
McNutt et al. \cite{mcnutt2020surfacing} adapt metamorphic software testing methods\cite{shin2023perceptual, segura2016metamorphicsurvey} to surface potential perceptual misreadings in a Vega-Lite visualization. 
In their work, they devise methods to verify that the visualization output is consistent against a range of inputs, \textit{e.g.,} shuffling the input data for a visualization and validating that the plotted data does not change. 
LitVis \cite{wood2019litvis} is an IDE plugin that provides live feedback to better align a visualization design against specification schemas. 
Draco \cite{DracoGitHub} is a visualization constraint representation system that can be used to auto-complete design guidelines\cite{moritz2018draco} and support the creation of automated visualization tools. 
Chen et al. created VizLinter \cite{chen2021vizlinter}, a tool that extends and refines Draco's constraint system to detect errors in existing visualizations and can suggest fixes or correct errors.
While Chen et al. suggest a potential use for VizLinter in educational settings, they do not discuss any uses of their tool in real-world visualization education scenarios.

Hopkins, et al. released VisuaLint \cite{hopkins2020visualint}, another linter for Vega-Lite visualizations that can display \textit{in-situ} errors within a visualization to aid developers in quickly discovering mistakes.
They also discuss a possibility of incorporating their tool for educational purposes but do not give concrete examples of this.
LitVis, VizLinter, and VisuaLint are limited to static analysis of visualizations and do not incorporate functionality that will analyze or evaluate the interactivity in a visualization.

However, D3 visualizations can incorporate rich interaction, including drag-and-drop, hover, filtering, etc. 
For this reason, these linters are not sufficient for observing and grading the changes that occur to a visualization when the user interacts and explores. 
Additionally, we do not find any tools that find mistakes or suggest corrections to D3 visualizations. 
One reason for this may be due to D3's lower level API that does not retain the same declarative characteristics that higher level tools offer, \textit{e.g.,} Vega-Lite, and requires deconstruction of the underlying components in order to analyze them.  

\subsubsection{Analyzing a D3 Visualization's Structure}
\label{sec:related_work:auto_grade_complex:analyze_struc}
Harper and Agrawala have produced work on deconstructing D3 visualizations so that they may be re-styled \cite{harper2014deconstructing}. 
Their approach entails traversing the Document Object Model (DOM) subtree of the visualization and recovering the data, marks, and internal mappings of a D3 visualization.  
They test for linear mappings between data and marks through use of linear regression or categorical mappings if linear data do not exist.
Hoque and Agrawala extend this approach with a search engine\cite{hoque2019searching} that leverages the deconstruction methods so that a user may search for the visual style of a D3 visualization.
The search engine tool is limited to only capturing linear mappings, such as marks encoded linearly on an axis and cannot capture complex layout generation, such as a node-link diagram using force-directed layout.

In our Data and Visual Analytics course\cite{CS6242DVA}, we expose students to a variety of visualization categories and therefore require a grading approach that has the capability to generalize beyond visualizations that use only linear vertical and horizontal axes.

Our auto-grading approach is inspired by Harper and Agrawala's D3 deconstruction methods \cite{harper2014deconstructing}.  
We further extend these methods and extract the components of many different types of visualizations, including those without pure horizontal and vertical axis layout, such as a visualization of geo-spatial data, a node-link diagram, or those that non-linear scaling, \text{e.g.}, logarithmic or square-root scales. 
Since D3 visualizations are browser-based \cite{2011-d3}, we leverage the cross-browser compatible Selenium WebDriver\cite{Selenium} software package for automated browser testing and adapt it to interact with visualizations, as a real user would.
This gap analysis motivates our the development of our auto-grader design and we present the respective design goals in the the next section.
\section{Design Challenges}
\label{sec:design_challenges}

Our goal is to build an auto-grader that scalably and precisely evaluates many types of D3 visualizations.  Based upon our 10 years of experience teaching and grading D3 visualizations and our survey of related work in Section~\ref{sec:related_work}, we identify four design challenges (\textbf{C1} - \textbf{C4}).

\begin{enumerate}[topsep=0mm, itemsep=1mm, parsep=1mm, leftmargin=6mm, label=\textbf{C\arabic*.}, ref=\textbf{C\arabic*}]

\item \label{challenge:feedback}
\textbf{Increased Iterations of Detailed Feedback.}
While D3 is  highly expressive in terms of what can be built, there is a steep learning curve for new D3 developers. \cite{mei2018design} Students may face implementation challenges\cite{battle2022exploring} and need multiple rounds of feedback as they work to complete their D3 visualization implementation. 
We find that they regularly engage the instructor staff on the class forum and during office hours to ask questions or request feedback while they develop their visualization.
There are limits to what feedback can be given since we do not have the resources to fairly or accurately ``pre-grade'' a submission. 
We need a systematic approach to frequently give students feedback on their progress and information on their submission grade that will decrease confusion and set clear expectations \cite{sherman2013impact,gramoli_mining_autograde}.

\item \label{challenge:scale}
\textbf{Scaling to Large Class Sizes.}
Our Data and Visual Analytics course has been simultaneously offered to on-campus undergraduate and graduate students, as well as to graduate students in two Online Master of Science degree programs in Computer Science\footnote{\url{https://omscs.gatech.edu}} and 
Analytics\footnote{\url{https://pe.gatech.edu/degrees/analytics}}
at Georgia Institute of Technology. 
As a result, we have increased enrollment from a few hundred students in the course's earlier offerings to more than 1,100 students in each recent semester \cite{CS6242DVA_enrollment}.  
At such a large scale, even a ``simple'' manual evaluating task taking only, say, 5 minutes to complete results in almost a hundred hours of total grading time \cite{maicus2020autograding}. 
Thus, we need a new, scalable approach to evaluate students' deliverables while offering the same depth and quality of instruction.

\item \label{challenge:variation}
\textbf{Accepting Variations in Visualization Designs.}
When we design a visualization assignment, the instructions include required data, visualization type, structure, and interactive features. 
We do not attempt an exhaustive description or expect that students will submit an implementation that exactly matches the examples we provide. 
This challenge requires implementing an auto-grader that is far more robust than the ``value/output-checking'' auto-graders\cite{Gao2016AutomatedFF,Parihar2017AutomaticGA} by accounting for variations between student submissions.

\item \label{challenge:interactivity}
\textbf{Grading Interactivity.}
D3 visualizations may be designed to incorporate a myriad of interactive features, ranging from simple actions such as \textit{hover} and \textit{click}, to more complex actions such as filtering a dataset, drag-and-drop, or displaying a subplot that is dynamically generated upon user selection.
Thus, our course has a key goal of teaching the design and development of richly interactive visualizations and to provide feedback and evaluation of those implementations.
However, existing techniques that detect mistakes were designed for static visualization components \cite{chen2021vizlinter, hopkins2020visualint} and manually grading every student's submission requires extensive time and is error prone \cite{maicus2020autograding}.
We need to develop a new automated approach that evaluates interactive features in a visualization that is scalable, being able to execute quickly for the large number of student submissions (\ref{challenge:scale}) and provide them with instant feedback.

\end{enumerate}

\begin{figure}
    \centering
    \includegraphics[width=\linewidth]{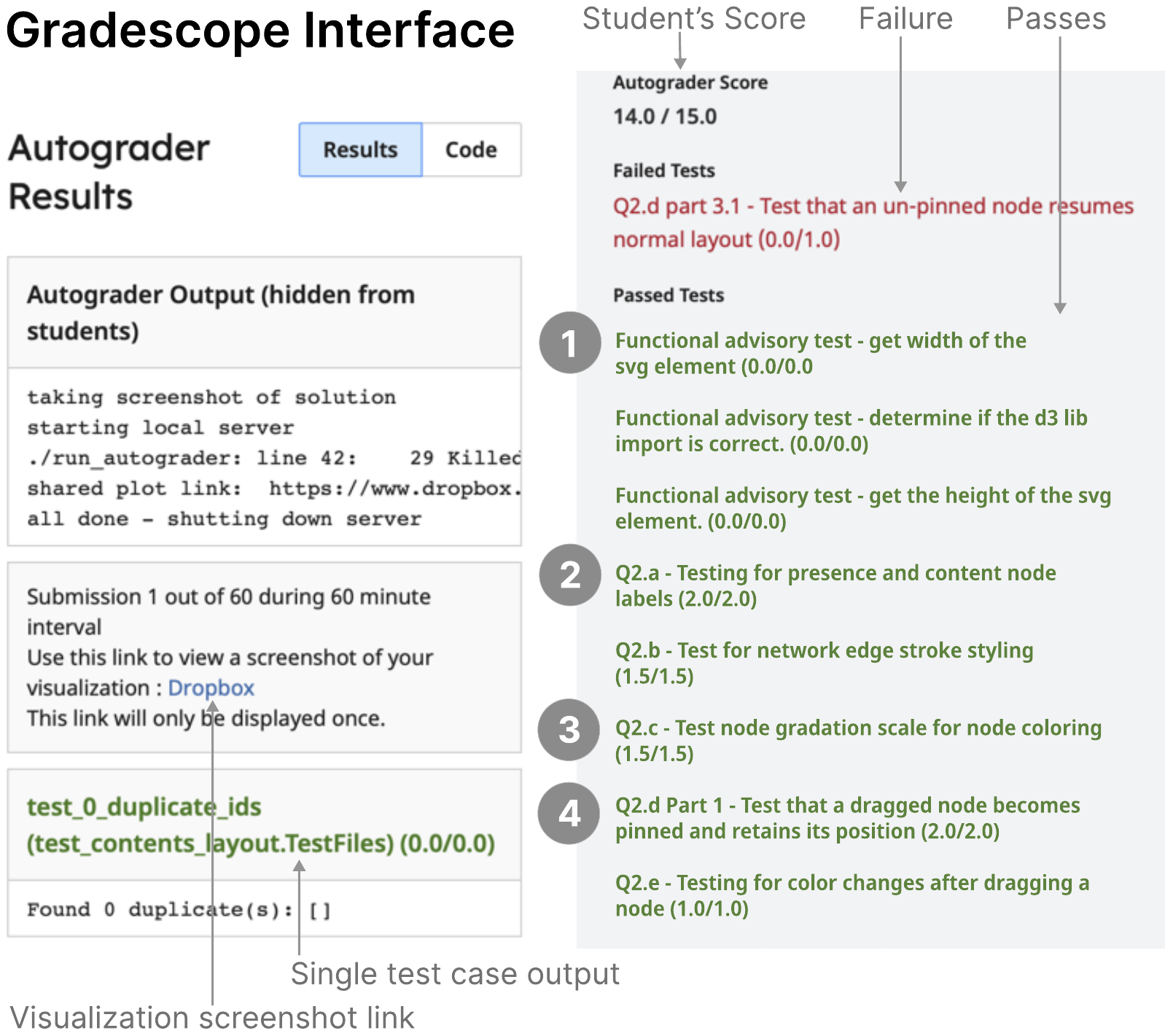}
    \caption{The Gradescope interface allows students to rapidly improve their D3 implementation by quickly giving feedback on four categories of test cases that evaluate the visualization based on an assignment rubric: (1) advisory on layout and required components; (2) mark appearance; (3) scale and data positioning; and (4) interactivity. Additionally, students may download a screenshot of their visualization to understand what the grader ``sees'' during grading.}
    \label{fig:gradescope_ui}
\end{figure} %
\begin{figure*}
    \includegraphics[width=\linewidth]{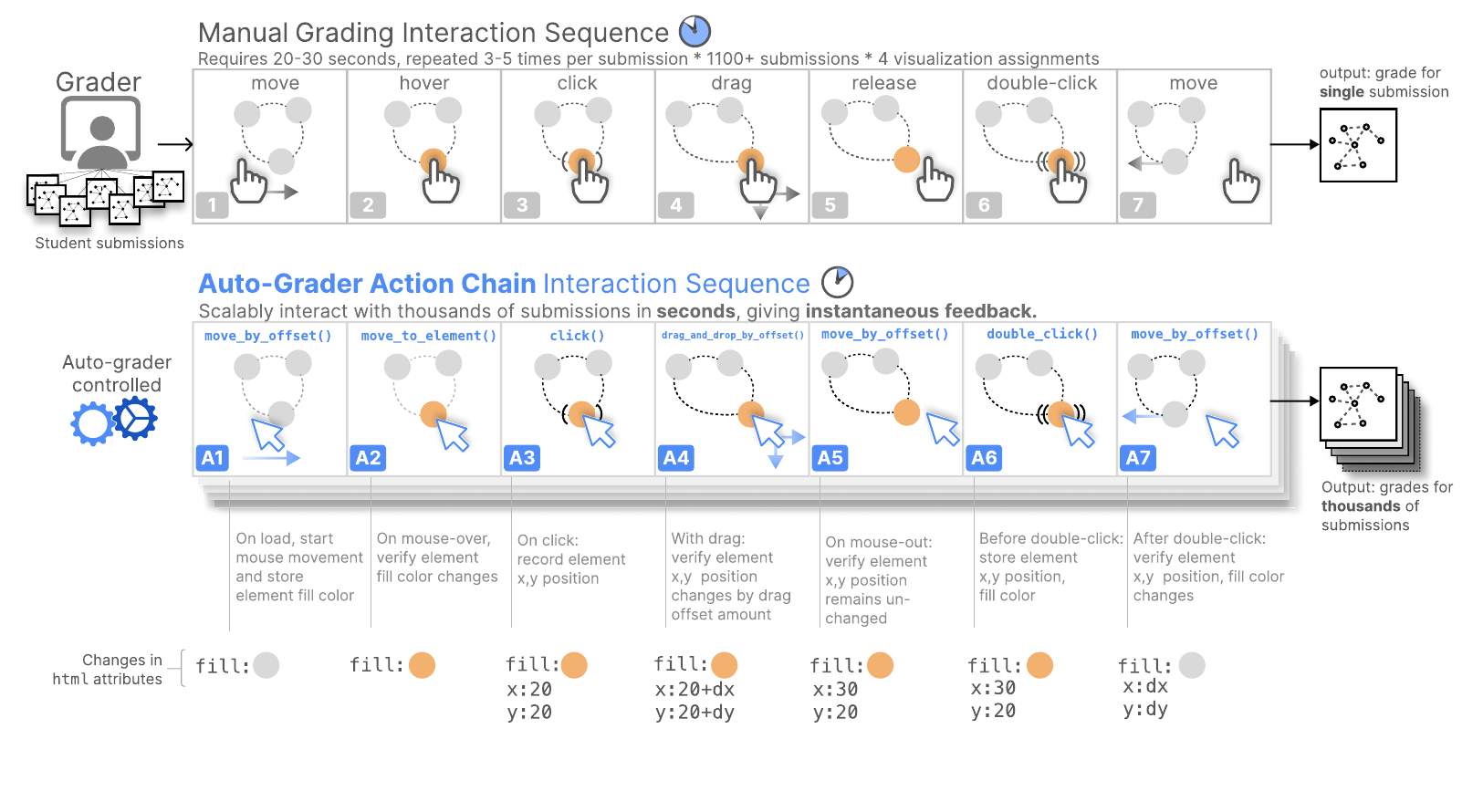}
    \centering
    \caption{Selenium Action Chains enable browser automation for complex tasks. \tool{} uses Action Chains to interact with a visualization by composing user-driven events, e.g.,  \inlinefig{9}{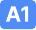} Moving the mouse across the visualization or  \inlinefig{9}{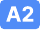} hovering over a data point.  Each interactive event drives a change in the DOM that is captured by our auto-grader.
    }
    \label{fig:action_chains}  
\end{figure*}

\section{\tool{}: Automatic D3 Grading}
\label{sec:design_goals}

\smallskip
Based on the identified design challenges (Sec. \ref{sec:design_challenges}), we distill the following key design goals (\textbf{G1-G4}) for \tool{}, an auto-grading method for D3 visualizations used in large scale instructional settings.

\begin{enumerate}[topsep=3mm, itemsep=1mm, parsep=1mm, leftmargin=6mm, label=\textbf{G\arabic*.}, ref=\textbf{G\arabic*}]

\item \label{goal:feedback}
\textbf{Rapid and frequent feedback interface.}
We use the Gradescope online assessment platform \cite{Gradescope} to host our D3 auto-graders.
This popular platform is used by over 2,600 universities for grading a variety of assignments: programming, quizzes, worksheets, and exams. 
Gradescope enables the instructional staff to configure programming assignments that allow students to submit their answers for an \textit{unlimited} number of times during the assignment window (Fig.~\ref{fig:gradescope_ui}), affording students with an opportunity to 
incrementally improve their design over time through several iterations of submission and viewing feedback (\ref{challenge:feedback})\cite{sherman2013impact}, e.g., a point score and textual description describing instances where they did not meet the assignment rubric. (Sec. \ref{sec:method:student_feedback})

\item \label{goal:distributed_auto_grading}
\textbf{Distributed auto-grading for thousands of students.}
Our goal is to scale instruction by further leveraging the capability of the Gradescope platform to support larger classes (\ref{challenge:scale}) for D3 assignments, thereby requiring fewer TAs to support and grade student submissions (Sec. \ref{sec:method:work_flow}) and allocating more TAs toward higher quality student interaction. (Sec. \ref{sec:evaluation:student_impact:improved_interact})

\item \label{goal:design_flexibility}
\textbf{Logic that preserves design flexibility.}
Our goal is to \textit{avoid} enforcing a single rigid solution that reduces the student's design autonomy. 
Instead, we encourage students to make their own design choices when building a D3 visualization with respect to colors, layout, chart size, visual encodings, etc., and flexibly grade (\ref{challenge:variation}) the student's implementation in a manner that does not penalize a submission that does not exactly match the example solution design(s). (Sec. \ref{sec:method:flexible_grading_design})

\item \label{goal:eval_interact}
\textbf{Evaluation of interactive visualizations.}
Our goal is to develop an auto-grader that evaluates rich interactivity and functions as a real user would (\ref{challenge:interactivity}). 
To accomplish this, the auto-grader must incorporate two high level functionalities (Sec. \ref{sec:method:eval_interaction}): 
\begin{itemize}
    \item Perform complex interactions, such as hovering over or clicking on a an element, drag-drop, filtering, scrolling, etc.
    \item Measure how the interaction effects the visualization, \textit{e.g.,} changes to the fill color, x/y position of an element.
\end{itemize}
\end{enumerate} %
\section{Auto-grading Method}
\label{sec:method}

In this section, we outline the implementation and function of \tool{} and demonstrate how it meets the four design goals (\textbf{G1} - \textbf{G4}).

\subsection{Student Feedback and Test Cases}
\label{sec:method:student_feedback}
\subsubsection{Student Submission Workflow}
\label{sec:method:student_feedback:submission_workflow}
For a D3 assignment, students upload visualization source code files to Gradescope and can generally view feedback on their submission within 30 seconds. 
Gradescope hosts the auto-grader and runs by rendering the visualization in a ``headless'' version of Google Chrome's web browser and scores the submission each time the student uploads a new version. 
The final grade is based on their highest scoring submission in an effort to encourage students to continue to improve their work without any imposed limit on submission frequency or running time limits.
When the auto-grader scores a student submission, it runs a suite of test cases that correspond to the rubric requirements and returns the total number of points earned along with textual feedback regarding which test cases passed or failed (Fig.~\ref{fig:gradescope_ui})
The auto-grader also captures an image of the visualization rendering and offers it to the student for download. 
This screenshot allows the student to evaluate how their visualization is ``seen'' by the auto-grader and could aid in critical error capture, \textit{e.g.}, a blank screenshot indicates that a critical error prevented the visualization from being rendered at all. 
\subsubsection{Test Case Categories}
\label{sec:method:student_feedback:test_case_categories}
In general, we use 4 categories 
of test cases for grading:

\begin{enumerate}[topsep=3mm, itemsep=1mm, parsep=1mm, leftmargin=6mm, label=\textbf{\arabic*.}]
\item \textbf{Advisory on Layout and Required Components} (Fig.~\ref{fig:gradescope_ui}.1).
\label{sec:method:student_feedback:test_case_categories:cat_1}
Advisory test cases are used to detect the visualization layout by extracting the dimensions, transformations, and margins as well as detecting the required components.  
Layout detection test cases form the assumptions that \tool{} uses for subsequent testing for axis scaling and data positioning.  
Required component test cases will alert the student if an element is missing, \textit{e.g.}, a missing x-axis. 
\item \textbf{Mark Appearance} (Fig.~\ref{fig:gradescope_ui}.2).
\label{sec:method:student_feedback:test_case_categories:cat_2}
Mark appearance test cases grade any required data sorting, sizing, and colors.
A data sorting test might evaluate if the bars in a bar chart are ordered by height, while a sizing test might check if the bars all contain a constant width.  
A color test could be used to ensure that all data for a certain grouping value are colored correctly (\textit{e.g.}, same color for all points on a scatterplot). 
\item \textbf{Scale and Data Positioning} (Fig.~\ref{fig:gradescope_ui}.3).
\label{sec:method:student_feedback:test_case_categories:cat_3}
Scale and data positioning test cases use the assumptions gathered in an advisory test case to grade the domain and range of axes and check that data are positioning correctly for an expected scaling function.
\item \textbf{Interactivity} (Fig.~\ref{fig:gradescope_ui}.4).
\label{sec:method:student_feedback:test_case_categories:cat_4}
Interactivity test cases perform interaction that results in a change to the DOM and then grade the result. 
This category is frequently implemented as a composition of interactivity testing and a subsequent mark appearance or scale and data positioning test.
\end{enumerate}

\subsection{Large Scale Auto-Grading Workflow}
\label{sec:method:work_flow}
We discuss our workflow for large-scale auto-grading and how this approach can be adopted for use in online and in-person offerings of a course, since most of the work for a D3 assignment is done outside of the classroom instructional periods.

\subsubsection{Easy Adoption}
\label{sec:method:work_flow:low_effort_adoption}
\tool{}'s architecture makes it easy for someone to adopt and implement their own D3 auto-grader, since the majority of the implementation involves writing unit tests, a typical skill for most computer science educators. 
Many of the test cases may be re-used with small modifications, thereby reducing the overhead on instructors writing test cases for a new visualization.
\textit{e.g.}, layout and required component test cases (Sec. \ref{sec:method:student_feedback:test_case_categories}.1) may be re-used since the specific required elements and layout information are abstracted into a rubric configuration YAML file (Sec. \ref{sec:deployment:viz_crs_assign_req}).  
Mark appearance test cases (Sec. \ref{sec:method:student_feedback:test_case_categories}.2) can be re-used with small modifications between similar designs, such as grading appearance of  \inlinefig{9}{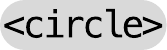} elements contained in (Fig. \ref{fig:example_charts}.1--3). In a similar manner, the axis scale test cases (\ref{sec:method:student_feedback:test_case_categories}.3) for the line chart shown in (Fig. \ref{fig:example_charts}.2) were used as a template for the line chart shown in (Fig. \ref{fig:example_charts}.3)
On average, we write 28 test cases per visualization using Python to capture the results and use Gradescope's API for score tabulation and formatting feedback. 

\subsubsection{Distributed Grading}
\label{sec:method:work_flow:wide_distribution}
\tool{} can be widely distributed and \textit{easily maintained} using Gradescope's integration with Docker. 
Each auto-grader is ``containerized'' and distributed as a Docker image and referenced by Gradescope from Dockerhub.
When an auto-grader needs revision, the modified source code is packaged by re-building the Docker image and pushing the new version to DockerHub.
Gradescope presents per-assignment dashboards for the instructor staff to see how many students have submitted and how many times the student has submitted on a particular question.  
In more difficult troubleshooting scenarios, the instructor can view specific error messages that are hidden from student's view to avoid divulging the solution or specific test case implementations.
Instructors can also establish a bash terminal session into the student's auto-grader session if needed. 
At the close of the submission period, the final grading can be exported from Gradescope system into a variety of supported Learning Management Systems (LMS), such as: Blackboard \cite{blackboard_lms}, Brightspace (D2L)\cite{brightspace_lms}, Canvas\cite{canvas_lms}, Moodle\cite{moodle_lms}, and Sakai\cite{sakai_lms}. 
This enables instructors to securely push grades directly to the LMS and prevents any errors that might occur when scoring by hand.  

\subsubsection{Parity between Development and Grading Environments}
\label{sec:method:work_flow:env_parity}
The auto-grader also helps prevent critical discrepancies that can exist between the student development environment and the grading environment that can affect the accurate scoring of a student submission.  
We ensure this by taking several steps to ensure that the student is developing their visualization using the same web browser version, libraries, and file structure that the auto-grader will use.
We also provide assignment templates that contain the same file directory structure that mirrors what the auto-grader will use with respect to the location of any data files and the D3-library references.
These measures assist in curtailing auto-grader tampering and reduce stress for the students as they are only required to submit the HTML, JavaScript, and CSS source files that they have edited. 

\subsection{Flexibly Grading Student Design}
\label{sec:method:flexible_grading_design}
We describe how \tool{} flexibly grades a student's design by capturing key attributes of a student's submission such as the chart size, layout, colors, and visual encodings (Fig. \ref{fig:diff_designs}.2) and evaluates the design with respect to the rubric. 
One of our key contributions is the novel use of adapting such existing browser-automation for visualization auto-grading, which has not been done before. 
Such adaptation goes beyond conventional ways of developing test cases for websites. For example, the test cases need to be carefully designed to allow for a great variety and combinations of student design choices, which is not typical for conventional test suites where test developers are able to pre-determine and control elements they wish to test for (\textit{e.g.}, position or scaling of a UI element derived from a requirement specification.) 
We also extend our test cases beyond basic fault detection to provide useful feedback that guides student improvement.

\subsubsection{Capturing Elements}
\label{sec:method:flexible_grading_design:elem_capture}
To initiate element capture, Harper and Agrawala's deconstruction method \cite{harper2014deconstructing} requires that the user click on the visualization and the subsequent event is captured through JavaScript code injection and initiates a crawl of the DOM sub-tree to find the respective chart elements.
Instead of intercepting a user's click event, we differ from this approach by  specifying a structure requirement following a predictable pattern convention found in many example D3 visualizations. We explicitly state in the assignment requirements (Sec. \ref{itm:DOM_struct}) which elements should contain certain \inlinefig{9}{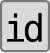} attributes or be placed within a \inlinefig{9}{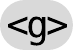} grouping element, the same approach that is used in this example from the official D3 documentation,\footnote{\url{https://observablehq.com/@d3/histogram\#Histogram}} all of the \inlinefig{9}{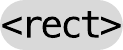} elements that represent bars in a bar chart are placed under one \inlinefig{9}{g-elem.pdf} element. 
For each D3 visualization, we present the students with a required chart structure and element ids that they must adopt in their implementation (Fig. \ref{fig:diff_designs}.2).  For a bar chart, a student would group all of the  \inlinefig{9}{rect-elem.pdf} elements representing bars in a group tag with an id of ``bars'' \inlinefig{9}{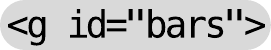}.  

During auto-grading, we use Selenium WebDriver's API\cite{SeleniumActionChain} to  locate and extract the HTML elements using this structure.
The Selenium WebDriver API stores all of the attributes and properties of the captured element as an object that can be queried during testing.
D3 visualization are typically comprised of scalable vector graphics (SVG) elements and can be mixed with HTML.  As a result, we can write a unit test that tests any SVG or HTML attribute, \textit{e.g.,} fill color, stroke color, x,y position, CSS classes.

\subsubsection{Capturing Scales, Axes}
\label{sec:method:flexible_grading_design:scale_axes}
During auto-grading, it is necessary to know if a datum is plotted correctly for a given scale type.
For example, if the visualization contains linearly scaled data, we need to know that datum's position is also scaled linearly.
The same also applies for non-linear displays of data, such as when a logarithmic scale is used.
Other visual encodings independent of Cartesian coordinate space could also be employed in a visualization, such as using a quantile scale to generate color gradations for a data mark.
Since D3 offers such a wide variety of scale types and methods for displaying chart axes, we need to evaluate both the scale domain and range, in order to evaluate how a mark should appear in a visualization. 

The student's choice for a D3 scale function call is not known to us, and we do not attempt to find it (\textit{i.e.}, we do not listen for API calls, use regex or parse JavaScript code).
For flexibly grading a scale used in a visualization, we must account for different design choices.
For example, a test case that checks for missing data points in a line chart (Fig.~\ref{fig:diff_designs}.1, design 2) would first identify the domain and range of the horizontal scale that the student uses (``year''). 
Then we match the range of our solution's scale with the student's, so that the test case can compute the expected positions of all data points (in student 2's visualization) to check for missing ones.
In our auto-graded D3 assignments, we specify which scale type that must be used but do not mandate any size dimensions or colors (Sec. \ref{sec:deployment:viz_crs_assign_req}).
Instead, we might instruct the student to:  ``use a linear scale with a domain of values that are represented in the provided data'', or ``use a quantile scale to generate the color scheme based on the average rating''.

\subsection{Performing and Grading Visualization Interaction}
\label{sec:method:eval_interaction}
Auto-grader flexibility is also extended to the interactive elements in a visualization and meets our goal of designing interactive data visualizations and promoting exploration of the presented data. 

We observe that interaction events within a browser-based visualization result in changes to the DOM.
If a student is building a visualization that implements a JavaScript function that intercepts a mouseover event and changes the fill color of some target element, then we can analyze the DOM structure and query the fill color of the target element before and after the mouseover event occurred to test if the interaction is correct.
The Selenium WebDriver API allows specification of user-actions that interact with a website in a natural way, supporting all interactions that a human user may perform. 
The same mouseover event is accomplished in Selenium by calling \texttt{move\_to\_element()}, a function that moves the mouse over a target element. 
Following the mouseover, we query the css attributes of a target element by calling \texttt{value\_of\_css\_property('fill')} without enforcing a specific color (Fig. \ref{fig:action_chains}-A2).
In a similar way, we can replicate click events by calling \texttt{click()} after moving the mouse over the desired element (Fig. \ref{fig:action_chains}-A3).
Some other interaction types include: hover, click, drag-and-drop, filtering, displaying pop-overs, hidden elements, sub-charts, and tool-tips.
Some interactions can only be accomplished by following a specific sequence of events. 
We use Selenium's \textit{Action Chains} \cite{SeleniumActionChain} API to compose sequences of interactions for arbitrary events, enabling the composition of rich interaction sequences that can be used to thoroughly interact with a visualization.
Fig. \ref{fig:action_chains} A1-A7 demonstrates a chain of events that are composed and used to evaluate that hovering over an object changes the fill color of an element, clicking on the element pins the node, dragging and dropping a node to a new position, and double-clicking un-pins the node. 
Beneath each action, we show the modifications to the element attribute that can be used for comparison in a unit test. %

\section{Large Scale D3 Auto-Grading Deployment}
\label{sec:deployment}

\subsection{Scale and Student Experience}
\label{sec:deployment:scale_student_exp}
\tool{} is the result of iterative design and development over two years by the instructional staff of Georgia Tech's \textit{CSE 6242 Data and Visual
Analytics}\footnote{\url{https://poloclub.github.io/\#cse6242}}. 
We presented our initial ideas for auto-grading a simple ``warm-up'' D3 exercise as a two-page non-archival research poster \cite{hull2021towards}.
Following the success of this deployment, we scaled our auto-graders to cover an entire homework assignment consisting of four D3 visualizations from different visualization categories.
Now, \tool{} auto-grades the entire D3 assignment consisting of four visualizations covering a variety of aspects of data processing, chart types, 
scale types, and interaction styles where students complete their work and have it graded entirely automatically using the Gradescope platform (Fig. \ref{fig:gradescope_ui})\cite{Gradescope}, without requiring any manual grading by the instructional staff. 
Currently, we have deployed our auto-grader for 4 semesters in Georgia Tech's CSE 6242: Data and Visual Analaytics course that has been offered for 10 years and was recently expanded with an online offering at the Master's level.
To date, more than 4,000 students have used \tool{}.

\begin{figure}
    \centering
    \includegraphics[width=0.95\linewidth]{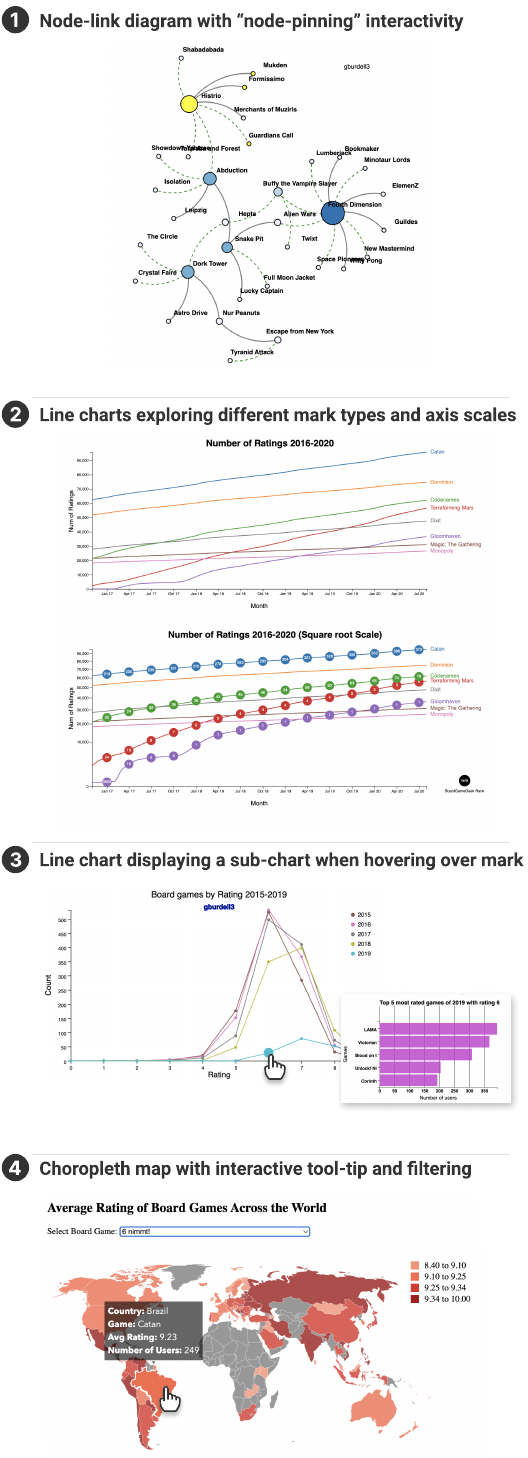}
    \caption{Example D3 Visualizations that students implement our auto-graded assignments.}
    \label{fig:example_charts}
\end{figure}

\subsection{Visualization Categories and Interaction}
\label{sec:deployment:viz_cat_interact}
Before the homework assignment is released, the students complete an introductory auto-graded D3 assignment consisting of a bar plot that introduces them to the D3 library and gives them practice with auto-graded assignments. The introductory assignment does not require interaction or require any data pre-processing.
Following this, the students receive a D3 homework assignment consisting of four visualizations, each covering a variety of aspects of data processing, chart types, scale types, interaction styles.
The following visualization types are used:

\begin{itemize} [topsep=2mm, itemsep=1mm, parsep=1mm, leftmargin=6mm]
    \item \textbf{Node-link diagram with ``node-pinning'' interactivity} (Fig.~\ref{fig:example_charts}.1). Nodes represent board games, and edges represent similarities between board games. Dragging a node ``pins'' (``freezes'') its position so that it will not be modified by the graph layout algorithm; double-clicking a pinned node unpins (unfreezes) its position, so it moves freely again.
    
    \item \textbf{Line charts exploring different mark types and axis scales} (Fig. \ref{fig:example_charts}.2). Lines and marks represent temporal patterns in the BoardGameGeek data to compare how the number of ratings grew. Symbols show board game ratings for a specific point in time. 
    
    \item \textbf{Line chart displaying sub-chart when user hovers over mark} (Fig. \ref{fig:example_charts}.3)
    Average board game ratings are displayed using an interactive frequency polygon line chart. A dynamically generated bar chart appears when the user hovers on a rating-count datum.
    
    \item \textbf{Choropleth Map} (Fig. \ref{fig:example_charts}.4) A Choropleth maps is used  to examine the popularity of different board games across the world. Selecting a different game from the drop-down menu causes the map to be re-colored. A pop-over tooltip menu displays detailed information when the user hovers over a country.
\end{itemize}
Our objective is to allow the student practice with building visualizations that allow the user to explore the data. 
To meet this goal, the  Force-Directed Graph, Line Chart, and Choropleth visualizations require the implementation of several user interactions, such as hover, click, drag-and-drop, and filtering using a drop-down menu.  

\subsection{Visualization Course Assignment Requirements}
\label{sec:deployment:viz_crs_assign_req}
The D3 homework assignment instructions are issued in a document that specifies \textit{rubrics} that describe each visualization and a \textit{DOM structure} that specifies the sequence and structure of required elements (Fig. \ref{fig:diff_designs}.2). 

\bigskip
\noindent
\textbf{Rubric items specify how the data should be displayed, or describe an interaction.}
For example, in one of our assignments where we ask students to create an interactive line chart (Fig. \ref{fig:example_charts}.3), we  provide the following rubric items (paraphrased for brevity):
\begin{itemize}
    \item \textit{``Create a line chart. Summarize the data by displaying the count of board games by rating for each year.''}
    \item \textit{``All axes should be automatically adjusted based on the data. Do not hard-code any values. Use a linear scale.''}
    \item \textit{``Create a horizontal bar chart, so that when hovering over a circle, that bar chart will be shown below the line chart.''}
    \item \textit{``The circle in the line chart should change to a larger size during mouseover to emphasize that it is the selected point and return to original size on mouseout.''}
\end{itemize}

\smallskip
\noindent
\textbf{The Document Object Model (DOM) structure is a lightweight, minimal specification}\label{itm:DOM_struct} that the visualization must adopt so that the auto grader can  locate the gradable elements (Fig. \ref{fig:diff_designs}.2). 
The structure is non-invasive, e.g., often times only adding an \inlinefig{9}{id_tag_elem.pdf} attribute to an svg element is adequate.  
This requirement does not restrict the student design choices or prevent them from adding additional elements.
The auto-grader developers encode the required DOM element structure in a rubric configuration file written in YAML. 
\textit{e.g.,} The required \inlinefig{9}{g-elem.pdf} groupings could be specified as 
\texttt{elements}, such as \texttt{g: [``lines'', ``circles'', ``legend'']}, signifying that the visualization requires a minimum of 3 group elements with the listed \inlinefig{9}{id_tag_elem.pdf} values.
When the auto-grader runs on the submission, the rubric configuration is referenced and we provide preliminary information to the student within Gradescope describing whether or not the required elements have been found in addition to the layout information (discussed in Sec. \ref{sec:method:flexible_grading_design:scale_axes}).
These advisory tests serve to assist the student in detecting critical implementation errors that would prevent the auto-grader from executing correctly and help the student align the structure of their submission. 
After verifying the structure and required elements, the remaining tests verify the rubric requirements and awards points for each test case that is passed.
We discuss the impact of this strategy in the next section. %
\section{Impact, Lessons Learned, and Limitations}
\label{sec:evaluation}

\bigskip
\subsection{Student Impact}
\label{sec:evaluation:student_impact}
\subsubsection{Receiving Instant Feedback to Improve Their Learning}
\label{sec:evaluation:student_impact:instant_feedback}
As a result of using Gradescope for submissions, students have a greatly simplified and transparent workflow as they complete work on their assignment. 
Students are able to see their scores immediately instead of waiting for the grading session to end.
This helped curtail a long-standing frustration that students often had when waiting for their scores or believing that they would have a higher score.  
Both Sherman et al.\cite{sherman2013impact} and Gramoli et al.\cite{gramoli_mining_autograde} also noted that students leverage instant feedback to improve their score on auto-graded programming assignments.

One example of improved learning is evidenced by the quantity of student questions on Ed Discussion\cite{eddiscussion}, the course’s discussion forum.
We compared the number of questions between the Fall 2020, the last semester before \tool{} was deployed and the most recent semester of Spring 2023 for the online section of the course.
Our experience in teaching this class over the years show that students are more likely to post assignment questions when they need clarification or assistance with an implementation while assignments that are considered less difficult consistently have lower question counts. 
After releasing \tool{} for the assignment described in (Sec. \ref{sec:deployment:viz_cat_interact}), we observed that the number of questions decreased from 1,965 posts to 1,356 posts (-31.0\%) for a nearly equivalent enrollment size: 920 in Fall 2020 and 926 in Spring 2023 (online sections), leading us to conclude that there is some reduction in confusion and improved learning experience.  
\begin{itemize}
    \item Node Link Diagram with ``node-pinning interactivity'' (Fig. \ref{fig:example_charts}.1), we observed a decrease from 360 posts to 260 posts (-27.8\%)
    \item Line charts exploring different mark types and axis scales (Fig. \ref{fig:example_charts}.2), we observed a decrease from 539 posts to 438 posts (-18.7\%)
    \item Line chart displaying sub-chart when hovering over mark (Fig. \ref{fig:example_charts}.3), we observed a decrease from 699 to 439 posts (-37.2\%)
    \item Choropleth map with tool-tip and filtering (Fig. \ref{fig:example_charts}.4), we observed a decrease from 367 to 219 posts (-40.3\%)
\end{itemize}

The feedback also helps improve the quality of the finished visualization by encouraging students to design their visualizations by using D3 principles and functions, instead of  hard-coding values. For example, as we have a test case that checks for the appropriate use of scale, for students who use for loops to create individual SVG elements (e.g., marks on a line chart) instead of using D3 selections and data join, they receive immediate feedback that calls attention to the need for using a scale.

\subsubsection{Facilitating Higher Quality Student Interaction}
\label{sec:evaluation:student_impact:improved_interact}
We also see evidence of improved learning through a re-distribution in types of student questions on Ed Discussion.
Specifically, by comparing the content of Ed Discussion posts across semesters, we found that the auto-grader had greatly reduced the number of repeated ``routine'' question posts related to grading (\textit{``would I lose points for \_\_\_?''), acceptable code functions (\textit{``can I use }\_\_\_?''}), and visualization styling. 
\tool{} preempted these types of questions through instant, informative feedback, allowing students to accomplish a significant amount of self-learning as they navigated these issues.  
We observed students completing more error discovery on their own and then helping others, allowing students to get answers at scale \cite{joyner2022teaching}. 
(\textit{``I experienced the above error... I did not set the value attribute of the option element (wrongly set a different attribute instead for the game name). Rectifying this fixed the issue for me.''}).
We are excited by having increased student-instructor interaction on more qualitative topics about gaining knowledge and experience with visualization design and implementation, and less on logistical issues that could distract students.
\subsubsection{Reduced Configuration Errors}
\label{sec:evaluation:student_impact:reduced_config_error}
Our approach has dramatically reduced configuration errors that students may experience, such as incorrectly referenced libraries or data files; before using auto-grader, such errors would prevent the visualization from being rendered or graded, and graders would need to manually resolved them. 
The auto-grader re-creates the student’s development environment and eliminates potential system configuration mismatch between student and instructor machines. 
As a result, \tool{}'s unified auto-grader environment eliminated the occurrences of re-grade requests for system configuration issues.

\subsubsection{Virtual Elimination of Re-Grade Requests}
\label{sec:evaluation:student_impact:eliminate_regrade}
Historically, we employed a re-grading process where students could appeal their grade if an error had been made. 
Large number of TA hours required to address this effort often resulted in multiple communications and some students submitted critiques that the process was frustrating due to small differences between the development environment and the TAs grading environment.
After implementing and deploying the auto-grader for more than 1,100 student in Fall 2021, the re-grade requests amount dropped 98\%, from more than 500 in the semester before to less than 10.  
Among these 10 re-grade requests, the most common error was that the student forgot to activate their highest scoring submission in Gradescope.

\subsection{Instructor Impact}
\label{sec:evaluation:instructor_impact}
\subsubsection{Reduced TA Grading Hours}
The auto-grader deployment resulted in a reduction of 400 TA hours (a conservative estimate) spent on grading effort every semester. Before using \tool{}, a grader would typically spend 5 minutes to grade one submission. For a class size of 1,100, this requires approximately 100 hours, or 5 part-time TAs per visualization.  Scaling to four D3 visualizations in one assignment required 400 hours, or 20 part-time TAs who focus only on grading.
This effort does not account for any hours spent actually interfacing with students by answering questions, conducting office hours, preparing future assignments, or reviewing re-grade requests.

\subsubsection{Virtual Elimination of Re-grade Work}
The reduction in re-grade requests (Sec.~\ref{sec:evaluation:student_impact:eliminate_regrade}) also resulted in a drastic reduction in the required TA hours needed to re-grade these assignments.
\tool{} also eliminated an entire class of re-grade requests caused by grading errors that arise from manual grading workflows, \textit{i.e., } we no longer require manual tabulation of scores or have to reconcile student and instructor development environment discrepancies.
\subsubsection{Increased Time for Qualitative Feedback}
In response to the reduction in required TA grading hours, we re-distributed TA effort to supporting the students questions during the assignment window through Office Hours and increased interaction on EdDiscussion. 
As discussed in the (Sec \ref{sec:evaluation:student_impact:improved_interact}), TAs were able to provide higher quality instruction and focus on more qualitative questions rather than repetitive questions on machine configuration, etc.
In addition, the instructional staff was able to spend more time collaborating and improving the auto-grader from semester to semester.

\subsection{Lessons Learned }
\label{sec:evaluation:lessons_learned}

\subsubsection{Balancing Focus Between structure and Design Choices}
\label{sec:evaluation:lessons_learned:balance_dom_design}
When we first started experimenting with auto-grading visualizations,
we required that the visualization follow a rigid DOM structure, allowing only small deviations, which unintentionally caused students to focus on getting the structure ``right'' rather than experimenting with different designs.
We quickly recognized that we could simplify the required structure (e.g., fewer elements in the DOM, and not requiring strict element ordering and nesting) without affecting the quality of the visualization. 
In hindsight, we realized that while
the DOM structure of a visualization is naturally produced during the implementation (\textit{e.g.}, it is natural to append \inlinefig{9}{rect-elem.pdf}  elements within a \inlinefig{9}{g-elem.pdf}  grouping element), 
it might not be natural for the students to consider it upfront when they begin their design process.
As we further refined our approach, we learned that we need to guide students toward a design structure that can be easily implemented and can provide sufficient structural features amenable to grading at a large scale.

\subsubsection{
Helping Students Learn Better Coding Practices}
\label{sec:evaluation:lessons_learned:coding_best_practice}
During periodic manual code reviews of student submissions, we learned that the most frequently missed test cases were due to students using a wide variety of methods to implement the specification of chart dimensions and margins.
D3 axes and scales depend upon this lay out, \textit{e.g.,} the x-axis size in pixels depends upon the chart width minus the left and right margins. 
We observed that when a student ``hard-codes'' the maximum width of the axis without referring to the chart dimensions they might make a small mistake \textit{e.g.,} even a few pixels can cause the plotted data point to have an incorrect horizontal positioning in the chart.  

To address this,  we suggested that students use the ``Margin Convention'' chart construction technique as a consistent way to layout a visualization. 
To help students use this technique, we designed additional advisory (no-penalty) test cases that present the student with a preliminary brief report of how the auto-grader detected their chart dimensions and margins.
The insight we share is that using \tool{} has elevated our teaching of D3 beyond just helping students just get the ``right'' answer but to also teach consistent coding conventions and train students to use a best practice.  
Using these techniques helps us address the frequent question: \textit{``What is the best implementation of this assignment?''} 
We find that teaching students to code using a best practice results in an improved implementation without necessarily showing them the exact code they should have written.

\subsubsection{
Designing Informative Test Cases}
We continuously monitor the discussion forums and office-hour conversations to obtain information about the students' experience.  
We learned that the limited feedback that only returns a generic index error, runtime error, or only displays limited information, \textit{e.g.}, ``var \texttt{a} and \texttt{b} are not equal'' are not helpful in a learning environment and caused frustration among students.
In response, we began designing test cases that provide multi-level feedback in an effort to help the student resolve complex programming issues.
Now, our unit testing design provides increased information about the rubric item being evaluated. 
For example, a test now produces:
\begin{enumerate}
\item Logging of helpful intermediate information during the test, alerting the student to any assumptions made by the auto-grader for the element being evaluated, \textit{e.g.}, ``Found a \texttt{<g>} element with id x-axis''.  This provides information that helps the students verify what the auto-grader has detected.
\item A display of actual versus expected results when the test case fails and common mistakes that help them troubleshoot. \textit{E.g.}, ``The x-axis was found in your visualization but the tick marks are not spaced at the 10-year interval. 
Found ticks [1980, 1985, \dots], but expected [1980, 1990, \dots]. See \url{https://d3js.org/d3-time#interval_every}''.
\end{enumerate}

\subsection{Limitations}
\label{sec:evaluation:limitations}
We discuss limitations of \tool{} to inform our plans for improvement and future work.
\subsubsection{Focus on Test Case and Score}
\label{sec:evaluation:limitations:test_case_focus}
The primary feedback within auto-grading is centered around the results of the suite of test cases used for the visualization.  
While we endeavor to provide adequate coverage and useful feedback, we acknowledge that there could also exist tendencies for the students to only work to get the maximum score and move on without further exploring learning in the design space.
Therefore, learning correlates with  the quality and coverage of the test cases used. We strive to
design test cases that provide learning opportunities that facilitate formative improvements and provide feedback that helps students under \textit{why} such improvements are beneficial (such as those discussed next in Sec. \ref{sec:evaluation:limitations:grade_design}).
\subsubsection{Grading Design of a Visualization} 
\label{sec:evaluation:limitations:grade_design}
In our course, we emphasize best design practices (e.g., color choices, chart layout, axis limits) but currently have limited test cases on color choices. 
If a student chose red and green colors without considering red-green color blindness, they would not currently receive feedback guiding them toward a more appropriate palette. 
This poses a question in how to best teach students in the auto-grading environment --- by providing advisory guidance or associating such a design choice with a point value?
We are planning an improvement that will capture additional considerations for color-blind safe palettes and proper contrast \cite{chartability}.

\subsubsection{Allowing Open-Ended Design}
\label{sec:evaluation:limitations:open_ended}
Our auto-graded D3 assignments are specified at a level that describes the type, appearance, and behavior of the visualization. 
However, we currently do not support grading open-ended implementations where the choice of visualization type or interactions is left to each student's decision.
For example, one student may choose to visualize temporal data using a bar chart while another chooses to make a line chart.
We recognize that this limitation could constrain students in exploring more approaches and that leads us to consider how extending \tool{} could permit this assignment type in the future.
\section{Discussion and Future Work}
\label{sec:future_work}
In this section, we share our observations on future work for extending \tool{} and its capabilities.
\subsection{Generalized Visualization Auto-Grading}
Because D3 is a base technology and originally contributed as a ``visualization kernel''\cite{2011-d3} it has been used as a foundation for  many other visualization toolkits/libraries: Vega, Vega-Lite, Observable Plot.
Since all of these libraries can produce SVG elements, their output can be easily accessed with and analyzed by reading the underlying DOM structure.
This poses a broader question on how \tool{} can be generalized to grade visualizations made with tools other than D3.
While \tool{} has not been used for grading other libraries, the method could easily be adapted to grade other visualizations that use HTML, CSS, and SVG. 
Adaptations of \tool{} could make it possible for instructors to teach different visualization technologies at a large-scale and auto-grade them using our method
\subsection{Incorporating Design Suggestions}
In the future, we want to increase focus on assisting students in design choices by helping students design charts that are accessible, \textit{e.g.}, colorblind-safe schemes, proper color contrast, and font-sizes. 
The Chartability\cite{chartability} project provides guidance on how to implement a set of auditing principles to promote visualization design that captures these ideals.
We share the view of Shin et al.\cite{shin2023perceptual} that the Chartability principles could be incorporated into an automated tool.
\tool{} could be extended to adopt a Chartability audit by adding test cases. 
We could extract the colors used for visual encodings, measure font size, or determine the color contrast by querying CSS property attributes and then display suggestions to the student about how they could improve their design, similar to the accessibility ``hints'' offered by WebHint \cite{webhint}.
We recommend that additional ``design'' unit tests should not be used to penalize the student but to teach them better design principles. 
In scenarios outside of teaching, \tool{} could be used to analyze design choices between different implementations of the same visualization.
\subsection{ML Methods for Visualization Interpretation and Design}
Auto-grading visualizations in general could generally benefit from machine learning methods that could be employed to extract and interpret content and assist in design \cite{wang2021survey} and even enable creation of more open-ended visualization assignments that give students more design autonomy and rely less upon prescriptive rubrics. 
Chen et al. use a multi-task deep neural network to extract visualization components from bitmap images to generate extensible templates \cite{chen2019auto_info_graphic_extract}.
Their work could be adapted extract information from visualizations that are not composed of web standards HTML, CSS, and SVG.
Hu et al. contributed VizML\cite{hu2019vizML}, an visualization recommendation system that uses ML to identify the top 5 design choices for a visualization based upon a specific dataset. 
DeepEye \cite{luo2018deepeye} uses supervised learning to recognize a ``good'' or ``bad'' visualizations based on expert rules and can also rank visualizations to determine which one is better. 
PlotThread \cite{tang2020plotthread} uses reinforcement learning to optimize the story line of a visualization and could also be use to improve or suggest changes to a student's work. 

\section{Conclusion}
\label{sec:conclusion}

In this work, we present \tool{}, an auto-grading method that enables scalable grading for the static and interactive components of.a D3 visualization. 
We discussed the key-features of our method and how it can be flexibly adapted to other visualization assignments. 
In our discussion, we highlight how \tool{} can be extended to support other visualization libraries or adapted to support incorporation of design suggestions and recommendations that will help students develop better visualizations.
We hope that our work helps students overcome the learning curve associated with learning a new visualization platform and that visualization courses that teach D3 can use our method or adapt it for other visualization platforms. 

\bibliographystyle{abbrv-doi-hyperref}

\bibliography{autograde-viz.bib}

\begin{thebibliography}{10}

\bibitem{CS444DataViz}
{CS444 - Data Visualization. University of Arizona}.
\newblock \url{https://cscheid.net/courses/fall-2019/csc444/}, 2019.
\newblock Accessed on: Mar 02, 2023.

\bibitem{CS171Viz}
{CS171 - Visualization. Harvard University}.
\newblock \url{https://www.cs171.org/2022/}, 2022.
\newblock Accessed on: March 2, 2023.

\bibitem{CS5630DatViz4DS}
{CS5630/CS6630 - Visualization for Data Science. The University of Utah}.
\newblock \url{https://www.dataviscourse.net/2022/syllabus/}, 2022.
\newblock Accessed on: Mar 02, 2023.

\bibitem{CSE512DataViz}
{CSE512 - Data Visualization. University of Washington}.
\newblock \url{https://courses.cs.washington.edu/courses/cse512/22sp/}, 2022.
\newblock Accessed on : Mar 02, 2023.

\bibitem{CS448DataViz}
{CS448B - Data Visualization. Stanford University}.
\newblock \url{https://online.stanford.edu/courses/cs448b-data-visualization},
  2023.
\newblock Accessed on: Mar 02, 2023.

\bibitem{CS6242DVA}
{CS6242 Data and Visual Analytics. Georgia Institute of Technology}.
\newblock \url{https://poloclub.github.io/cse6242-2023spring-online/}, 2023.
\newblock Accessed on: Jan 08, 2023.

\bibitem{eddiscussion}
{EdDiscussion}.
\newblock \url{https://edstem.org}, 2023.
\newblock Accessed on: Aug, 08, 2022.

\bibitem{CS6242DVA_enrollment}
{Enrollment history of CS6242 Data and Visual Analytics. Georgia Institute of
  Technology}.
\newblock \url{https://poloclub.github.io/\#cse6242}, 2023.
\newblock Accessed on: Jan 08, 2023.

\bibitem{Ahmed2020CharacterizingTP}
U.~Z. Ahmed, N.~Srivastava, R.~Sindhgatta, and A.~Karkare.
\newblock Characterizing the pedagogical benefits of adaptive feedback for
  compilation errors by novice programmers.
\newblock In {\em Proc. ACM/IEEE 42nd International Conference on Software
  Engineering: Software Engineering Education and Training}, ICSE-SEET '20, p.
  139–150. ACM, New York, 2020.
  \href{https://doi.org/10.1145/3377814.3381703}
{doi: {{%
10\hspace{.1pt}\discretionary{.}{%
}{.}\hspace{.4pt}1145\discretionary{/}{%
}{/}3377814\hspace{.1pt}\discretionary{.}{%
}{.}\hspace{.4pt}3381703}}}


\bibitem{battle2022exploring}
L.~Battle, D.~Feng, and K.~Webber.
\newblock Exploring d3 implementation challenges on stack overflow.
\newblock In {\em 2022 IEEE Visualization and Visual Analytics (VIS)}, pp.
  1--5, 2022. \href{https://doi.org/10.1109/VIS54862.2022.00009}
{doi: {{%
10\hspace{.1pt}\discretionary{.}{%
}{.}\hspace{.4pt}1109\discretionary{/}{%
}{/}VIS54862\hspace{.1pt}\discretionary{.}{%
}{.}\hspace{.4pt}2022\hspace{.1pt}\discretionary{.}{%
}{.}\hspace{.4pt}00009}}}


\bibitem{blackboard_lms}
{Blackboard Inc.}
\newblock {Blackboard LMS}.
\newblock
  \url{https://www.blackboard.com/en-mea/teaching-learning/learning-management},
  2023.
\newblock Accessed on: Mar 01, 2023.

\bibitem{2011-d3}
M.~Bostock, V.~Ogievetsky, and J.~Heer.
\newblock D³ data-driven documents.
\newblock {\em IEEE Trans. Vis. Comput. Graph.}, 17(12):2301--2309, 2011.
  \href{https://doi.org/10.1109/TVCG.2011.185}
{doi: {{%
10\hspace{.1pt}\discretionary{.}{%
}{.}\hspace{.4pt}1109\discretionary{/}{%
}{/}TVCG\hspace{.1pt}\discretionary{.}{%
}{.}\hspace{.4pt}2011\hspace{.1pt}\discretionary{.}{%
}{.}\hspace{.4pt}185}}}


\bibitem{chen2021vizlinter}
Q.~Chen, F.~Sun, X.~Xu, Z.~Chen, J.~Wang, and N.~Cao.
\newblock Vizlinter: A linter and fixer framework for data visualization.
\newblock {\em IEEE Trans. Vis. Comput. Graph.}, 28(1):206--216, 2022.
  \href{https://doi.org/10.1109/TVCG.2021.3114804}
{doi: {{%
10\hspace{.1pt}\discretionary{.}{%
}{.}\hspace{.4pt}1109\discretionary{/}{%
}{/}TVCG\hspace{.1pt}\discretionary{.}{%
}{.}\hspace{.4pt}2021\hspace{.1pt}\discretionary{.}{%
}{.}\hspace{.4pt}3114804}}}


\bibitem{chen2019auto_info_graphic_extract}
Z.~Chen, Y.~Wang, Q.~Wang, Y.~Wang, and H.~Qu.
\newblock Towards automated infographic design: Deep learning-based
  auto-extraction of extensible timeline.
\newblock {\em IEEE Trans. Vis. Comput. Graph.}, 26(1):917--926, 2020.
  \href{https://doi.org/10.1109/TVCG.2019.2934810}
{doi: {{%
10\hspace{.1pt}\discretionary{.}{%
}{.}\hspace{.4pt}1109\discretionary{/}{%
}{/}TVCG\hspace{.1pt}\discretionary{.}{%
}{.}\hspace{.4pt}2019\hspace{.1pt}\discretionary{.}{%
}{.}\hspace{.4pt}2934810}}}


\bibitem{brightspace_lms}
{D2L Corporation}.
\newblock {Brightspace LMS}.
\newblock \url{https://www.d2l.com/brightspace/}, 2023.
\newblock Accessed on: Mar 01, 2023.

\bibitem{chartability}
F.~Elavsky.
\newblock Chartability.
\newblock \url{https://chartability.github.io/POUR-CAF/}, 2021.
\newblock Accessed on: Jul 1, 2021.

\bibitem{Gao2016AutomatedFF}
J.~Gao, B.~Pang, and S.~S. Lumetta.
\newblock Automated feedback framework for introductory programming courses.
\newblock In {\em Proc. ACM Conference on Innovation and Technology in Computer
  Science Education}, ITiCSE '16, p. 53–58. ACM, New York, 2016.
  \href{https://doi.org/10.1145/2899415.2899440}
{doi: {{%
10\hspace{.1pt}\discretionary{.}{%
}{.}\hspace{.4pt}1145\discretionary{/}{%
}{/}2899415\hspace{.1pt}\discretionary{.}{%
}{.}\hspace{.4pt}2899440}}}


\bibitem{Sen2005DARTDA}
P.~Godefroid, N.~Klarlund, and K.~Sen.
\newblock Dart: Directed automated random testing.
\newblock {\em SIGPLAN Not.}, 40(6):213–223, 2005.
  \href{https://doi.org/10.1145/1064978.1065036}
{doi: {{%
10\hspace{.1pt}\discretionary{.}{%
}{.}\hspace{.4pt}1145\discretionary{/}{%
}{/}1064978\hspace{.1pt}\discretionary{.}{%
}{.}\hspace{.4pt}1065036}}}


\bibitem{Gradescope}
{Gradescope Inc.}
\newblock Gradescope.
\newblock \url{https://www.gradescope.com}, 2020.
\newblock Accessed on: May 01, 2020.

\bibitem{gramoli_mining_autograde}
V.~Gramoli, M.~Charleston, B.~Jeffries, I.~Koprinska, M.~McGrane, A.~Radu,
  A.~Viglas, and K.~Yacef.
\newblock Mining autograding data in computer science education.
\newblock In {\em Proc. Australasian Computer Science Week Multiconference},
  number~1 in ACSW '16, pp. 1--10. ACM, New York, 2016.
  \href{https://doi.org/10.1145/2843043.2843070}
{doi: {{%
10\hspace{.1pt}\discretionary{.}{%
}{.}\hspace{.4pt}1145\discretionary{/}{%
}{/}2843043\hspace{.1pt}\discretionary{.}{%
}{.}\hspace{.4pt}2843070}}}


\bibitem{Gulwani2014FeedbackGF}
S.~Gulwani, I.~Radi\v{c}ek, and F.~Zuleger.
\newblock Feedback generation for performance problems in introductory
  programming assignments.
\newblock In {\em Proc. ACM SIGSOFT International Symposium on Foundations of
  Software Engineering}, FSE 2014, p. 41–51. ACM, New York, 2014.
  \href{https://doi.org/10.1145/2635868.2635912}
{doi: {{%
10\hspace{.1pt}\discretionary{.}{%
}{.}\hspace{.4pt}1145\discretionary{/}{%
}{/}2635868\hspace{.1pt}\discretionary{.}{%
}{.}\hspace{.4pt}2635912}}}


\bibitem{harper2014deconstructing}
J.~Harper and M.~Agrawala.
\newblock Deconstructing and restyling d3 visualizations.
\newblock UIST '14, p. 253–262. ACM, New York, 2014.
  \href{https://doi.org/10.1145/2642918.2647411}
{doi: {{%
10\hspace{.1pt}\discretionary{.}{%
}{.}\hspace{.4pt}1145\discretionary{/}{%
}{/}2642918\hspace{.1pt}\discretionary{.}{%
}{.}\hspace{.4pt}2647411}}}


\bibitem{hopkins2020visualint}
A.~K. Hopkins, M.~Correll, and A.~Satyanarayan.
\newblock Visualint: Sketchy in situ annotations of chart construction errors.
\newblock In {\em Computer Graphics Forum}, vol.~39, pp. 219--228. Wiley Online
  Library, 2020. \href{https://doi.org/10.1111/cgf.13975}
{doi: {{%
10\hspace{.1pt}\discretionary{.}{%
}{.}\hspace{.4pt}1111\discretionary{/}{%
}{/}cgf\hspace{.1pt}\discretionary{.}{%
}{.}\hspace{.4pt}13975}}}


\bibitem{hoque2019searching}
E.~Hoque and M.~Agrawala.
\newblock Searching the visual style and structure of d3 visualizations.
\newblock {\em IEEE Trans. Vis. Comput. Graph.}, 26(1):1236--1245, 2020.
  \href{https://doi.org/10.1109/TVCG.2019.2934431}
{doi: {{%
10\hspace{.1pt}\discretionary{.}{%
}{.}\hspace{.4pt}1109\discretionary{/}{%
}{/}TVCG\hspace{.1pt}\discretionary{.}{%
}{.}\hspace{.4pt}2019\hspace{.1pt}\discretionary{.}{%
}{.}\hspace{.4pt}2934431}}}


\bibitem{hu2019vizML}
K.~Hu, M.~A. Bakker, S.~Li, T.~Kraska, and C.~Hidalgo.
\newblock Vizml: A machine learning approach to visualization recommendation.
\newblock In {\em Proc. CHI Conference on Human Factors in Computing Systems},
  CHI '19, p. 1–12. ACM, New York, 2019.
  \href{https://doi.org/10.1145/3290605.3300358}
{doi: {{%
10\hspace{.1pt}\discretionary{.}{%
}{.}\hspace{.4pt}1145\discretionary{/}{%
}{/}3290605\hspace{.1pt}\discretionary{.}{%
}{.}\hspace{.4pt}3300358}}}


\bibitem{hull2021towards}
M.~Hull, C.~Guerin, J.~Chen, S.~Routray, and D.~H. Chau.
\newblock Towards automatic grading of d3.js visualizations.
\newblock {\em arXiv preprint arXiv:2110.11227}, 2021.
  \href{https://doi.org/10.48550/arXiv.2110.11227}
{doi: {{%
10\hspace{.1pt}\discretionary{.}{%
}{.}\hspace{.4pt}48550\discretionary{/}{%
}{/}arXiv\hspace{.1pt}\discretionary{.}{%
}{.}\hspace{.4pt}2110\hspace{.1pt}\discretionary{.}{%
}{.}\hspace{.4pt}11227}}}


\bibitem{canvas_lms}
{Instructure Inc.}
\newblock {Canvas LMS}.
\newblock \url{https://www.instructure.com/canvas}, 2023.
\newblock Accessed on: Mar 01, 2023.

\bibitem{joyner2022teaching}
D.~Joyner.
\newblock {\em Teaching at Scale: Improving Access, Outcomes, and Impact
  Through Digital Instruction}.
\newblock Taylor \& Francis, 2022. \href{https://doi.org/10.4324/9781003274834}
{doi: {{%
10\hspace{.1pt}\discretionary{.}{%
}{.}\hspace{.4pt}4324\discretionary{/}{%
}{/}9781003274834}}}


\bibitem{Lo2019eLearnVizTools}
L.~Y.-H. Lo, Y.~Ming, and H.~Qu.
\newblock Learning vis tools: Teaching data visualization tutorials.
\newblock In {\em 2019 IEEE Visualization Conference (VIS)}, pp. 11--15, 2019.
  \href{https://doi.org/10.1109/VISUAL.2019.8933751}
{doi: {{%
10\hspace{.1pt}\discretionary{.}{%
}{.}\hspace{.4pt}1109\discretionary{/}{%
}{/}VISUAL\hspace{.1pt}\discretionary{.}{%
}{.}\hspace{.4pt}2019\hspace{.1pt}\discretionary{.}{%
}{.}\hspace{.4pt}8933751}}}


\bibitem{luo2018deepeye}
Y.~Luo, X.~Qin, N.~Tang, and G.~Li.
\newblock Deepeye: Towards automatic data visualization.
\newblock In {\em 2018 IEEE 34th International Conference on Data Engineering
  (ICDE)}, pp. 101--112, 2018. \href{https://doi.org/10.1109/ICDE.2018.00019}
{doi: {{%
10\hspace{.1pt}\discretionary{.}{%
}{.}\hspace{.4pt}1109\discretionary{/}{%
}{/}ICDE\hspace{.1pt}\discretionary{.}{%
}{.}\hspace{.4pt}2018\hspace{.1pt}\discretionary{.}{%
}{.}\hspace{.4pt}00019}}}


\bibitem{maicus2020autograding}
E.~Maicus, M.~Peveler, A.~Aikens, and B.~Cutler.
\newblock Autograding interactive computer graphics applications.
\newblock SIGCSE '20, p. 1145–1151. ACM, New York, 2020.
  \href{https://doi.org/10.1145/3328778.3366954}
{doi: {{%
10\hspace{.1pt}\discretionary{.}{%
}{.}\hspace{.4pt}1145\discretionary{/}{%
}{/}3328778\hspace{.1pt}\discretionary{.}{%
}{.}\hspace{.4pt}3366954}}}


\bibitem{mcnutt2020surfacing}
A.~McNutt, G.~Kindlmann, and M.~Correll.
\newblock Surfacing visualization mirages.
\newblock CHI '20. ACM, New York, 2020.
  \href{https://doi.org/10.1145/3313831.3376420}
{doi: {{%
10\hspace{.1pt}\discretionary{.}{%
}{.}\hspace{.4pt}1145\discretionary{/}{%
}{/}3313831\hspace{.1pt}\discretionary{.}{%
}{.}\hspace{.4pt}3376420}}}


\bibitem{mei2018design}
H.~Mei, Y.~Ma, Y.~Wei, and W.~Chen.
\newblock The design space of construction tools for information visualization:
  A survey.
\newblock {\em J. Vis. Lang. Comput.}, 44:120--132, 2018.
  \href{https://doi.org/10.1016/j.jvlc.2017.10.001}
{doi: {{%
10\hspace{.1pt}\discretionary{.}{%
}{.}\hspace{.4pt}1016\discretionary{/}{%
}{/}j\hspace{.1pt}\discretionary{.}{%
}{.}\hspace{.4pt}jvlc\hspace{.1pt}\discretionary{.}{%
}{.}\hspace{.4pt}2017\hspace{.1pt}\discretionary{.}{%
}{.}\hspace{.4pt}10\hspace{.1pt}\discretionary{.}{%
}{.}\hspace{.4pt}001}}}


\bibitem{moodle_lms}
{Moodle PTY Ltd.}
\newblock {Moodle LMS}.
\newblock \url{https://moodle.com/solutions/lms/}, 2023.
\newblock Accessed on: Mar 01, 2023.

\bibitem{DracoGitHub}
D.~Moritz.
\newblock {Draco: Formalizing Visualization Design Knowledge as Constraints}.
\newblock \url{https://uwdata.github.io/draco/}, 2018.
\newblock Accessed on: Mar 09, 2023.

\bibitem{moritz2018draco}
D.~Moritz, C.~Wang, G.~L. Nelson, H.~Lin, A.~M. Smith, B.~Howe, and J.~Heer.
\newblock Formalizing visualization design knowledge as constraints: Actionable
  and extensible models in draco.
\newblock {\em IEEE Trans. Vis. Comput. Graph.}, 25(1):438--448, 2019.
  \href{https://doi.org/10.1109/TVCG.2018.2865240}
{doi: {{%
10\hspace{.1pt}\discretionary{.}{%
}{.}\hspace{.4pt}1109\discretionary{/}{%
}{/}TVCG\hspace{.1pt}\discretionary{.}{%
}{.}\hspace{.4pt}2018\hspace{.1pt}\discretionary{.}{%
}{.}\hspace{.4pt}2865240}}}


\bibitem{webhint}
{Open JS Foundation}.
\newblock {WebHint}.
\newblock \url{https://https://webhint.io}, 2023.
\newblock Accessed on: Jun 29, 2023.

\bibitem{Parihar2017AutomaticGA}
S.~Parihar, Z.~Dadachanji, P.~K. Singh, R.~Das, A.~Karkare, and
  A.~Bhattacharya.
\newblock Automatic grading and feedback using program repair for introductory
  programming courses.
\newblock In {\em Proc. ACM Conference on Innovation and Technology in Computer
  Science Education}, ITiCSE '17, p. 92–97. ACM, New York, 2017.
  \href{https://doi.org/10.1145/3059009.3059026}
{doi: {{%
10\hspace{.1pt}\discretionary{.}{%
}{.}\hspace{.4pt}1145\discretionary{/}{%
}{/}3059009\hspace{.1pt}\discretionary{.}{%
}{.}\hspace{.4pt}3059026}}}


\bibitem{2017-vega-lite}
A.~Satyanarayan, D.~Moritz, K.~Wongsuphasawat, and J.~Heer.
\newblock Vega-lite: A grammar of interactive graphics.
\newblock {\em IEEE Trans. Vis. Comput. Graph.}, 23(1):341--350, 2017.
  \href{https://doi.org/10.1109/TVCG.2016.2599030}
{doi: {{%
10\hspace{.1pt}\discretionary{.}{%
}{.}\hspace{.4pt}1109\discretionary{/}{%
}{/}TVCG\hspace{.1pt}\discretionary{.}{%
}{.}\hspace{.4pt}2016\hspace{.1pt}\discretionary{.}{%
}{.}\hspace{.4pt}2599030}}}


\bibitem{segura2016metamorphicsurvey}
S.~Segura, G.~Fraser, A.~B. Sanchez, and A.~Ruiz-Cortés.
\newblock A survey on metamorphic testing.
\newblock {\em IEEE Trans. Softw. Eng.}, 42(9):805--824, 2016.
  \href{https://doi.org/10.1109/TSE.2016.2532875}
{doi: {{%
10\hspace{.1pt}\discretionary{.}{%
}{.}\hspace{.4pt}1109\discretionary{/}{%
}{/}TSE\hspace{.1pt}\discretionary{.}{%
}{.}\hspace{.4pt}2016\hspace{.1pt}\discretionary{.}{%
}{.}\hspace{.4pt}2532875}}}


\bibitem{sakai_lms}
{SGW Communications}.
\newblock {Sakai LMS}.
\newblock \url{https://www.sakailms.org}, 2023.
\newblock Accessed on: Mar 01, 2023.

\bibitem{sherman2013impact}
M.~Sherman, S.~Bassil, D.~Lipman, N.~Tuck, and F.~Martin.
\newblock Impact of auto-grading on an introductory computing course.
\newblock {\em J. Comput. Sci. Coll.}, 28(6):69–75, 2013.
  \href{https://dl.acm.org/doi/10.5555/2460156.2460171}
{doi: {{%
doi\discretionary{/}{%
}{/}10\hspace{.1pt}\discretionary{.}{%
}{.}\hspace{.4pt}5555\discretionary{/}{%
}{/}2460156\hspace{.1pt}\discretionary{.}{%
}{.}\hspace{.4pt}2460171}}}


\bibitem{shin2023perceptual}
S.~Shin, S.~Hong, and N.~Elmqvist.
\newblock Perceptual pat: A virtual human system for iterative visualization
  design.
\newblock {\em arXiv preprint arXiv:2303.06537}, 2023.
  \href{https://doi.org/10.48550/arXiv.2303.06537}
{doi: {{%
10\hspace{.1pt}\discretionary{.}{%
}{.}\hspace{.4pt}48550\discretionary{/}{%
}{/}arXiv\hspace{.1pt}\discretionary{.}{%
}{.}\hspace{.4pt}2303\hspace{.1pt}\discretionary{.}{%
}{.}\hspace{.4pt}06537}}}


\bibitem{Selenium}
{Software Freedom Conservancy}.
\newblock Selenium.
\newblock \url{https://www.selenium.dev}, 2021.
\newblock Accessed on: February 01, 2021.

\bibitem{SeleniumActionChain}
{Software Freedom Conservancy}.
\newblock {Selenium Action Chain}.
\newblock
  \url{https://www.selenium.dev/selenium/docs/api/py/webdriver/selenium.webdriver.common.action_chains.html},
  2021.
\newblock Accessed on: March 14, 2021.

\bibitem{tang2020plotthread}
T.~Tang, R.~Li, X.~Wu, S.~Liu, J.~Knittel, S.~Koch, T.~Ertl, L.~Yu, P.~Ren, and
  Y.~Wu.
\newblock Plotthread: Creating expressive storyline visualizations using
  reinforcement learning.
\newblock {\em IEEE Trans. Vis. Comput. Graph.}, 27(2):294--303, 2021.
  \href{https://doi.org/10.1109/TVCG.2020.3030467}
{doi: {{%
10\hspace{.1pt}\discretionary{.}{%
}{.}\hspace{.4pt}1109\discretionary{/}{%
}{/}TVCG\hspace{.1pt}\discretionary{.}{%
}{.}\hspace{.4pt}2020\hspace{.1pt}\discretionary{.}{%
}{.}\hspace{.4pt}3030467}}}


\bibitem{wang2021survey}
Q.~Wang, Z.~Chen, Y.~Wang, and H.~Qu.
\newblock A survey on ml4vis: Applying machine learning advances to data
  visualization.
\newblock {\em IEEE Trans. Vis. Comput. Graph.}, 28(12):5134--5153, 2022.
  \href{https://doi.org/10.1109/TVCG.2021.3106142}
{doi: {{%
10\hspace{.1pt}\discretionary{.}{%
}{.}\hspace{.4pt}1109\discretionary{/}{%
}{/}TVCG\hspace{.1pt}\discretionary{.}{%
}{.}\hspace{.4pt}2021\hspace{.1pt}\discretionary{.}{%
}{.}\hspace{.4pt}3106142}}}


\bibitem{wilcox2015role}
C.~Wilcox.
\newblock The role of automation in undergraduate computer science education.
\newblock SIGCSE '15, p. 90–95. ACM, New York, 2015.
  \href{https://doi.org/10.1145/2676723.2677226}
{doi: {{%
10\hspace{.1pt}\discretionary{.}{%
}{.}\hspace{.4pt}1145\discretionary{/}{%
}{/}2676723\hspace{.1pt}\discretionary{.}{%
}{.}\hspace{.4pt}2677226}}}


\bibitem{wood2019litvis}
J.~Wood, A.~Kachkaev, and J.~Dykes.
\newblock Design exposition with literate visualization.
\newblock {\em IEEE Trans. Vis. Comput. Graph.}, 25(1):759--768, 2019.
  \href{https://doi.org/10.1109/TVCG.2018.2864836}
{doi: {{%
10\hspace{.1pt}\discretionary{.}{%
}{.}\hspace{.4pt}1109\discretionary{/}{%
}{/}TVCG\hspace{.1pt}\discretionary{.}{%
}{.}\hspace{.4pt}2018\hspace{.1pt}\discretionary{.}{%
}{.}\hspace{.4pt}2864836}}}


\end{thebibliography}

\end{document}